\documentclass[12pt]{article}

\catcode`\@=11

\def\section{\@startsection{section}{1}{\z@}{3.5ex plus 1ex minus
 .2ex}{2.3ex plus .2ex}{\bf}}

\def\thesubsection{\arabic{section}.\arabic{subsection}}
\renewcommand{\subsection}[1]{\addtocounter{subsection}{1}
\vspace{2.5mm}\par\noindent {\it \thesubsection . #1}\par
 \vspace{0.5mm} }
\catcode`\@=12
\newfont{\mbm}{msbm10 scaled\magstep1}

\def\be{\begin{equation}}
\def\ee{\end{equation}}
\def\ba{\begin{eqnarray}}
\def\ea{\end{eqnarray}}

\begin{document}
\begin{titlepage}

\rightline{CPTH-PC 089.1000}
\rightline{ROM2F-2000/33}
\rightline{{hep-th/0010279}}
\vskip 2cm
\centerline{{\large\bf Type-I vacua and brane transmutation}}
\vskip 1cm
\centerline{Carlo Angelantonj${}^{a}$ and 
Augusto Sagnotti${}^{b}$}
\vskip 0.5cm
\centerline{\it ${}^a$ Centre de Physique Th\'eorique de l'\'Ecole
Polytechnique\footnote{Unit{\'e} mixte du CNRS et de l'EP, UMR 7644.},
F-91128 Palaiseau}
\vskip 0.3cm
\centerline{\it ${}^b$ Dipartimento di Fisica, Universit\'a di Roma
``Tor Vergata'' e INFN, Sezione di Roma 2}
\centerline{\it Via della Ricerca Scientifica 1, I-00133 Rome}
\vskip  1.0cm
\begin{abstract}
In a classic paper, Fradkin and Tseytlin showed how magnetic deformations
can be introduced
in open strings. In this contribution we review some recent work on 
type-I vacua with magnetised branes and describe the role 
of additional discrete deformations, related to quantised values of the 
NS-NS antisymmetric tensor $B_{ab}$.
\end{abstract}
\vskip 36pt
\begin{center}
{\it 
Contribution to the Conference on ``Quantisation, Gauge Theory, and 
Strings'' \vskip 10 pt 
dedicated to the memory of Professor Efim S. Fradkin}
\end{center}
\end{titlepage}

\section{Introduction}

We are pleased and honoured to contribute to this celebration of 
Prof. Efim S. Fradkin. His work has had a profound influence on String 
Theory, and in particular on the recent results that we have 
chosen to review here.
In addition, one of us (A.S.) had the privilege of meeting Prof. Fradkin
on several occasions, where anyone could enjoy his friendly attitude in 
sharing his insights, while witnessing his pervasive enthusiasm for
scientific research.

Our starting point is the classic paper \cite{ft}, where Fradkin and 
Tseytlin showed how to introduce magnetic deformations in open 
strings and linked the Born-Infeld action to String Theory.
These results, supplemented by the elegant canonical analysis
of \cite{acny}, are central to most current developments related to
D-brane physics and non-commutative geometry. Our interest 
in the magnetic deformations of \cite{ft}, however, is not directly 
related to non-commutative geometry. Rather, it can be traced to the
proposal of Bachas \cite{costas} of attaining the breaking of space-time
supersymmetry via string compactifications on magnetised tori. 
This setting can be regarded as a string realisation of the early field theory 
work of Witten \cite{wito32} and, as in that case, was actually restricted
to configurations with vanishing instanton density.

In a recent work \cite{aads}, we relaxed the restriction to vanishing 
instanton density,
and showed how the brane content of toroidal and orbifold models 
can be affected in an interesting way. Magnetised D9 branes
can then acquire a D5 charge, and special configurations 
with self-dual internal fields can even result in 
new interesting vacua with unbroken {\it supersymmetry}. Whereas the
phenomenon of brane transmutation, a consequence of the peculiar
Wess-Zumino coupling of D-branes, was previously discussed in \cite{ghm},
together with the restrictions associated with unbroken supersymmetry 
\cite{bdl},
the models presented in \cite{aads} and reviewed here are the first
consistent vacuum configurations for type-I strings where this setting is
realised. 
They stand out for their gauge groups with peculiar ranks and for the
structure of their matter representations, that occur in multiple families.
This construction actually requires a small additional step 
beyond \cite{costas}, the 
extension of magnetic deformations to orbifold models. 
The reason for this is somewhat technical: a supersymmetric 
configuration requires that only
D9 and D5 branes be present, as opposed to their antibranes, and only 
orbifolds, as opposed to tori, can contain the O5 planes capable of 
absorbing the D5 charge of magnetised D9 branes. 
The resulting constructions may be regarded as explicit string 
realisations of the inverse of the process originally
advocated in \cite{wsi}, where a five-brane is recovered from
small instantons: reverting this process one can associate
conventional fat instantons to magnetised D9 branes\footnote{We are 
grateful to A. Uranga for a discussion on this point.}. 

We shall conclude with some new results on the 
effect of a quantised NS-NS field $B_{ab}$ on these magnetised branes. 
As usual, this field is projected out in conventional type-I vacua \cite{cargese} 
that, however, allow interesting discrete deformations associated 
with its discrete values.
In tori \cite{bps}, these result in gauge groups of reduced rank and allow 
one to
interpolate between the orthogonal and symplectic cases. In orbifolds 
\cite{kst,bsb2},
these effects are accompanied by a multiplet structure
for the fixed points, that results in the appearance of several tensor
multiplets in the projected closed sector and of multiple families
of 95 states, as originally noted in the rational construction of \cite{bs}.
Although we 
shall confine our attention to the D9-D5 case, T-duality can be 
used to relate this setting to a number of similar ones involving other
types of branes. This applies, in particular, to the discrete deformations
associated to $B_{ab}$, that can be related to discrete choices for 
the geometry of the brane configuration, as in \cite{abgk}.

\section{Magnetised orbifolds and brane transmutation}

Let us begin by reviewing briefly the results of \cite{aads}.
Some intuitive field theory arguments suffice to expose the essence
of the phenomenon, and are well captured by
the low-energy effective action for D9 branes in an internal Abelian
background \footnote{As in \cite{aads}, the (dimensionless) magnetic fields 
differ from the conventional 
ones by a $2 \pi \alpha'$ rescaling.},
\be
{\cal S}_9 =  - T_{(9)} \int_{{\cal M}_{10}} \!\!\!\!\! {\rm d}^{10} x \; 
e^{-\phi} \sum_{a=1}^{32} \; \sqrt{-
\det \left( g_{10} + q_a F \right)} - \mu_{(9)}
\sum_{p,a} \;
\int_{{\cal M}_{10}} \!\!\! e^{q_a F} \wedge C_{p+1} + \ldots \ \ ,
\label{s9}
\ee
where $a$ labels the types of Chan-Paton (CP) charges that couple
to the magnetic fields with strength $q_a$, and
\be
T_{(p)} = \sqrt{\pi \over 2\kappa^2 } \; \left(
2 \pi  \sqrt{\alpha'} \right)
^{3-p} = | \mu_{(p)} | \ .
\ee
Here $T$ and $\mu$ are
the tension and the R-R charge for a type-I D$p$ brane \cite{pol}, while
$\kappa$ defines the ten-dimensional Newton constant 
$G_N^{(10)}=\kappa^2/8 \pi$.
To illustrate the phenomenon, it suffices to consider the
geometry ${\cal M}_{10} = {\cal M}_6 \times T^2 \times T^2$ with
constant Abelian magnetic fields $H_1$ and $H_2$ 
lying in the two internal tori.
These are monopole fields, and thus satisfy the Dirac
quantisation conditions
\be
q \, H_i \, v_i = k_i \ \qquad (i=1,2) \ , \label{dirac}
\ee
where, aside from powers of
$2 \pi$, $v_i=R_i^{(1)} R_i^{(2)}/\alpha'$ are the dimensionless 
volumes of the two tori of radii $R_i^{(1)}$
and $R_i^{(2)}$, $k_i$ are the degeneracies of the 
corresponding Landau levels and $q$ is the elementary electric charge
for the system. As
anticipated, we forego the restriction in \cite{wito32,costas} 
and actually pick a pair of Abelian fields aligned with
the same U(1) subgroup. This affects the Wess-Zumino coupling, giving
rise to an effective D5 charge, so that
\ba
{\cal S}_9 &=& - \; T_{(9)} \int_{{\cal M}_{10}}\!\!\!\!\!  {\rm d}^{10} x \; 
e^{-\phi} \; \sqrt{-g_6} \; \sum_{a=1}^{32} \; 
\sqrt{(1 +  q_a^2 H_1^2 ) 
(1 + q_a^2 H_2^2 )} \nonumber \\
& & - \; 32 \, \mu_{(9)} 
\int_{{\cal M}_{10}} C_{10} \; - \; \left(2 \pi \sqrt{\alpha'}\right)^4 
\; \mu_{(9)}
\; v_1 v_2 \; H_1 \; H_2\; 
\sum_{a=1}^{32} \; q_a^2 \; \int_{{\cal M}_{6}}
C_6   \ , \label{s9special}
\ea
where ${g_6}$ denotes the six-dimensional 
space-time metric, and where for simplicity
we have chosen an identity metric in the internal space. 
In particular, if the two internal fields have identical magnitudes, 
for the resulting (anti)self-dual configuration the action becomes
\ba
{\cal S}_9 &=&  - \; 32 \int_{{\cal M}_{10}} \!\!\! 
\left( {\rm d}^{10} x  \; \sqrt{-g_6}\; T_{(9)}\;
e^{-\phi} + \mu_{(9)} \; C_{10} \right) \nonumber \\
& &- \sum_{a=1}^{32} \left(\frac{q_{a}}{q}\right)^2 \; 
\int_{{\cal M}_{6}} \left( {\rm d}^{6} x \;  \sqrt{-g_6}\; 
|k_1 k_2| \; T_{(5)} \; e^{-\phi} + k_1 k_2 \; \mu_{(5)} \; 
C_{6} \right) \ . \label{s9fin}
\ea
Therefore, not only the Dirac quantisation conditions (\ref{dirac}) 
have compensated the integration over the internal tori, but in
the second line of (\ref{s9fin})
the additional powers of $\alpha'$ have nicely converted
$T_{(9)}$ and $\mu_{(9)}$ into $T_{(5)}$ and $\mu_{(5)}$.
As a result, a D9 brane on a magnetised $T^2 \times T^2$ indeed
mimics a D5 brane or a D5 antibrane according to whether the
orientations of $H_1$ and $H_2$, reflected by the relative sign
of $k_1$ and $k_2$, are identical or opposite.

We can now extend the analysis to String Theory, following \cite{cargese}. A
precise control of the CFT exhibits very nicely several properties of 
the effective action, including new couplings to twisted states of
orbifold models. As we anticipated, supersymmetry is generically 
broken, but supersymmetric vacuum configurations 
can be obtained starting
from an orbifold that normally requires the introduction of D5
branes. The simplest such instance is the six-dimensional
compactification on
$(T^2 \times T^2)/Z_2$ with Klein-bottle projection
\begin{equation}
{\cal K} = {\textstyle{1\over 4}} \left\{ (Q_o + Q_v) (0;0) \left[ P_1 P_2 +
W_1 W_2 \right] + 16\times 2 (Q_s + Q_c ) (0;0) \left( {\eta \over
\vartheta_4 (0)} \right)^2 \right\} \ , \label{klein}
\end{equation}
that corresponds to the introduction of ${\rm O}9_+$ and ${\rm O}5_+$ planes,
and thus to a projected ${\cal N}= (1,0)$ supersymmetric closed spectrum 
with one tensor multiplet and 20 hypermultiplets. 
For the sake of brevity, in this Section 
we shall confine our attention to the models of \cite{aads} without D5 branes, 
where the O5 charge is fully compensated by the magnetised 
D9 branes. The characters used in eq. (\ref{klein}) and in
the following are in general combinations of theta-functions with 
non-vanishing
arguments, that extend the standard $(1,0)$ supersymmetric combinations, 
and are described in detail in \cite{aads}.

As in our field
theory considerations, we introduce a pair of internal
magnetic fields aligned with the same U(1) subgroup of SO(32), and
we restrict our attention to the maximal residual gauge group,
${\rm U}(m) \times {\rm U}(n)$, with $m + n = 16$.
In writing the direct-channel annulus amplitude, let us 
begin by recalling \cite{acny} that a uniform
magnetic field with components $H_1$ and $H_2$ in the two internal tori 
alters the boundary conditions for open strings,
shifting their mode frequencies by
\be
z_i^{\rm L,R} \; = \; \frac{1}{\pi} \; \left[ \; \tan^{-1}( q_{{\rm L}}\,
H_i) \; + \; 
\tan^{-1}( q_{{\rm R}}\, H_i) \; \right] \ ,
\ee
where $q_{{\rm L}}$ ($q_{{\rm R}}$) denotes the charge of the left (right)
end of the open string with respect to the U(1) field $H_i$.  
A further novelty \cite{acny} 
is displayed by
``dipole'' strings, with opposite end charges, whose
oscillator modes are unaffected, but whose world-sheet coordinates undergo
a complex ``boost'', so that their Kaluza-Klein momenta $m_i$ are 
rescaled according to
\be
m_i \ \to \ \frac{m_i}{\sqrt{1 \; + \; q_a^2 H_i^2}}  \quad . \label{boost}
\ee

The direct-channel annulus amplitude is
\begin{eqnarray}
{\cal A} &=& {\textstyle{1\over 4}} \Biggl\{ (Q_o + Q_v)(0;0) \left[
(m+\bar m)^2 P_1 P_2 + 2 n \bar{n} \tilde P_1 \tilde P_2 \right]
\nonumber
\\
&-& 2 (m+\bar m) (n + \bar{n}) (Q_o + Q_v )(z_1 \tau ; z_2 \tau
) {k_1 \eta \over
\vartheta_1 (z_1 \tau)} {k_2 \eta \over \vartheta_1 (z_2 \tau)} 
\nonumber
\\
&-& ( n^2 + \bar{n}^2 ) (Q_o + Q_v ) (2 z_1 \tau ; 2 z_2 \tau ) 
{2 k_1 \eta \over
\vartheta_1 (2 z_1 \tau)} {2 k_2 \eta \over \vartheta_1 (2 z_2 \tau)} 
\nonumber 
\\
&-& \left[ (m-\bar m)^2 -2 n\bar n \right] (Q_o - Q_v ) (0;0)
\left( {2\eta \over \vartheta_2 (0)}\right)^2 
\nonumber
\\
&-& 2 (m-\bar m) (n - \bar{n}) (Q_o - Q_v ) (z_1 \tau ; z_2 \tau) 
{2\eta \over \vartheta_2
(z_1 \tau)} {2\eta \over \vartheta_2 (z_2 \tau)} 
\nonumber
\\
&-& (n^2 + \bar{n}^2) (Q_o - Q_v ) (2z_1 \tau ; 2z_2 \tau)
{2\eta \over \vartheta_2
(2z_1 \tau)} {2\eta \over \vartheta_2 (2z_2 \tau)} \Biggr\} \ , 
\end{eqnarray}
while the corresponding M\"obius amplitude is
\begin{eqnarray}
{\cal M} &=& -{\textstyle{1\over 4}} \Biggl\{ 
(\hat Q_o + \hat Q_v )(0;0) \left[ (m+\bar m) P_1 P_2  \right]
\nonumber
\\
&-& ( n + \bar{n}) (\hat Q_o + \hat Q_v ) (2z_1 \tau ; 2z_2 \tau) {2 k_1
\hat\eta \over \hat \vartheta_1 (2z_1\tau)} {2 k_2
\hat\eta \over \hat \vartheta_1 (2z_2\tau)}
\nonumber
\\
&-& \left( m+ \bar m \right) (\hat Q_o - \hat Q_v )(0;0) \left(
{2\hat\eta \over \hat \vartheta_2 (0)}\right)^2 \label{mobsusy}
\\
&-& (n + \bar{n}) (\hat Q_o - \hat Q_v ) (2 z_1 \tau ; 2 z_2 \tau )
{2\hat\eta \over \hat\vartheta_2 (2z_1\tau)}
{2\hat\eta \over \hat\vartheta_2 (2z_2\tau)} \Biggr\} \ .
\nonumber
\end{eqnarray}
The arguments $z_i$  ($2z_i$) are associated to strings with one
(two) charged ends while, for the sake of brevity, both the imaginary modulus 
$\frac{1}{2} i t$ of ${\cal A}$ and the complex modulus $
\frac{1}{2} + \frac{1}{2} i t$ of ${\cal M}$ are denoted
by the same symbol $\tau$, although
the proper ``hatted'' contributions to the M\"obius amplitude are explicitly
indicated. $P_i$ and $W_i$ are conventional momentum
and winding sums for the two-tori, while a ``tilde'' denotes a sum with
momenta ``boosted'' as in (\ref{boost}). Finally, 
$m$ and $n$ (together with their conjugates $\bar{m}$ and $\bar{n}$) 
are CP multiplicities for the D9 brane, and several terms with
with opposite $z_i$ arguments
have been grouped together, using the symmetries of 
the Jacobi theta-functions.

For generic magnetic fields, the open spectrum is indeed non-super\-sym\-metric
and develops Nielsen-Olesen instabilities \cite{no}. As emphasised in
\cite{costas}, the emergence of these tachyonic modes can be ascribed to the
magnetic couplings of the internal components of gauge fields. 
For instance, small magnetic fields
affect the mass formula for the untwisted string modes according to
\be
\Delta M^2 \; = \; \frac{1}{2 \pi \alpha'} \; \sum_{i=1,2} 
\left[  (2 n_i + 1) |(q_{\rm L} + q_{\rm R}) H_i| + 
2 (q_{\rm L} + q_{\rm R}) \Sigma_i H_i \right] \ ,
\ee
where the first term originates from the Landau levels and
the second from the magnetic moments of the spins $\Sigma_i$. 
For the internal
components of the vectors, the magnetic moment coupling generally
overrides the zero-point contribution, leading to tachyonic modes,
unless $|H_1|=|H_2|$,
while for spin-$\frac{1}{2}$ modes it can at most compensate it. 
Moreover, if $H_1=H_2$ the supersymmetry charge, 
that belongs to $C_4 C_4$, is also unaffected. 
On the other hand, if $H_1= - H_2$ one obtains models with ``brane
supersymmetry breaking'', similar in spirit to those of \cite{bsb,bsb2,
bsb3,bsb4,bsb5}.
However, in this case supersymmetry is broken
on the whole magnetised D9 brane, since we are working effectively in
the presence of blown-up instantons.

The untwisted R-R tadpole conditions arising from the $C_4 S_2 C_2$ 
sector read
\be
\left[ m+\bar m + n + \bar{n} - 32 +  q^2  H_1 H_2 (n +
\bar n ) \right] \sqrt{v_1 v_2} 
 - {32 \over \sqrt{v_1 v_2}}   =0 \ , \label{rrutad}
\ee
aside from terms that vanish after identifying the multiplicities
of conjugate representations $(m,\bar{m})$ and $(n,\bar{n})$.
The additional (untwisted) R-R tadpole conditions 
from $Q_o$ and $Q_v$ are compatible with (\ref{rrutad}) and do not add further 
constraints. This expression reflects the familiar Wess-Zumino
coupling of eq. (\ref{s9}), and therefore the various powers of $H$
correspond to R-R forms of different degrees. 
In particular, as we anticipated in our field theory discussion,
the term bilinear in the magnetic fields has a very
neat effect: it charges the D9 brane with respect to the
six-form potential. This can be seen very clearly making use of the
quantisation condition  (\ref{dirac}), that turns the tadpole
conditions (\ref{rrutad}) into
\begin{eqnarray}
m+\bar m + n + \bar n &=& 32 \ ,
\nonumber
\\
k_1 k_2 (n + \bar n )  &=& 32 \ . \label{urrt}
\end{eqnarray}

In a similar fashion, the untwisted NS-NS tadpoles exhibit very nicely 
their relation to the 
Born-Infeld term in (\ref{s9}), and can be linked to its derivatives
with respect to 
the corresponding moduli, while the twisted NS-NS tadpoles display new
couplings to twisted states present in the effective Lagrangian.  
A novelty is that some of the tadpoles are {\it not} perfect
squares, as a result of the peculiar behaviour of the internal magnetic
deformations under world-sheet time reversal. 
All these features are described in some detail in \cite{aads}.
While for generic internal magnetic fields it is impossible to satisfy
the NS-NS tadpoles, the supersymmetric choice $H_1=H_2$ 
makes them all nicely compatible with the R-R ones.

We can now describe the low-lying spectrum of a model without 
D5 branes and with
$k_1 = k_2 =2$, the minimal Landau-level degeneracies that on this 
$Z_2$ orbifold are compatible with the positivity of the direct channel. 
Although the closed spectrum is the standard one,
and comprises the ${\cal N}=(1,0)$ 
gravitational multiplet, together with one tensor multiplet and
twenty hypermultiplets, the open spectrum is quite different from
the familiar one of \cite{bs,gp}, with gauge group 
${\rm U}(16)|_9 \times {\rm U}(16)|_5$.
Having excluded the D5 branes, the solution of the
tadpole conditions yields $m=12$, $n=4$, and the result
is a rather unusual supersymmetric $Z_2$ model, with a gauge group of rank 16, 
${\rm U}(12) \times {\rm U}(4)$, and with charged 
hypermultiplets in the representations $(66 + \overline{66},1)$,
in five copies of the $(1,6 + \overline{6})$, and in four copies of
the $(\overline{12},4)$.
A distinctive feature of this spectrum, that is free of all irreducible
gauge and gravitational anomalies, consistently with the vanishing of
all RR tadpole conditions \cite{pc}, as well as of the similar ones
in \cite{aads}, is that some of the matter occurs in multiple families. 
This peculiar phenomenon is a consequence of the multiplicities of
Landau levels, that in these $Z_2$ orbifolds with vanishing $B_{ab}$
are multiples of two for each magnetised torus. Notice that, 
when D5 branes are also present \cite{aads}, one is led in general
to rank reductions, but not simply by powers of two 
as in the presence of a quantised $B_{ab}$ \cite{bps,kst,bsb2}.
These are not the first
concrete examples of brane transmutation in type I vacua but, to
the best of our knowledge, they are the first supersymmetric ones. Indeed,
$Z_2$ orientifolds without D5 branes appeared previously in
\cite{abg}, where magnetised fractional D9 branes were used to build
six-dimensional asymmetric orientifolds with ``brane
supersymmetry breaking''. 

\section{Introducing a quantised $B_{ab}$}

In this Section we extend the construction of \cite{aads} 
to allow for quantised values
of the NS-NS antisymmetric tensor $B_{ab}$, whose rank will be
denoted by $r$.

As discussed in \cite{kst,bsb2}, the quantised $B_{ab}$ has a twofold
effect on the Klein-bottle amplitude. The winding lattice now 
involves a projector, just like the
transverse annulus amplitude of the toroidal model discussed in \cite{bps}.
Moreover, as in \cite{bsb2} the $\Omega$ eigenvalues of some of the twisted 
contributions are reverted, as demanded by the transverse channel
amplitude, whose
coefficients are to be perfect squares. Thus
\begin{eqnarray}
{\cal K} &=& {\textstyle {1\over 4}} (Q_o + Q_v ) (0;0) \left[ P_1 P_2 +
2^{-4} \sum_\epsilon W_1 W_2 e^{{2 i \pi \over \alpha '} n^{{\rm T}} B
\epsilon} \right] \nonumber
\\
&& + {2^{(4-r)/2}\over 2} (Q_s + Q_c ) (0;0) \left(
{\eta \over \vartheta_4 (0) }\right)^2  \ . \label{dkleinhb}
\end{eqnarray}

Turning to the open sector, for the sake of brevity we shall again confine
our attention to models without D5 branes, since the other cases can be
easily reconstructed from these results. The quantised $B_{ab}$ has
a twofold effect on ${\cal A}$: it affects the momentum lattice and
endows the contributions related to the Landau levels with additional
multiplicities depending on the rank $r$ of $B_{ab}$. Thus
\begin{eqnarray}
{\cal A} &=& {\textstyle{1\over 4}} \Biggl\{ (Q_o + Q_v ) (0;0) \Biggl[
(m+\bar m)^2 2^{r-4} \sum_\epsilon P_1 (B) P_2 (B) 
\nonumber
\\
& & \qquad\qquad\qquad + 2 n \bar n
2^{r-4} \sum_\epsilon \tilde P _1 (B) \tilde P _2 (B) \Biggr]
\nonumber
\\
&& - 2 \cdot 2^r (m+\bar m) (n+\bar n ) (Q_o + Q_v) (z_1 \tau ; z_2 \tau)
{k_1 \eta \over \vartheta_1 (z_1 \tau)}
{k_2 \eta \over \vartheta_1 (z_2 \tau)} \nonumber
\\ 
&&- 2^r (n^2 + \bar n^2 ) (Q_o + Q_v) (2 z_1 \tau ; 2 z_2 \tau ) 
{2 k_1 \eta \over \vartheta_1 (2 z_1 \tau)}
{2 k_2 \eta \over \vartheta_1 (2 z_2 \tau)} \nonumber
\\
&&- \left[ (m- \bar m )^2 - 2 n \bar n \right] (Q_o - Q_v ) (0;0)
\left( {2 \eta \over \vartheta_2 (0)}\right)^2 \nonumber
\\
&& - 2 (m-\bar m)(n-\bar n) (Q_o - Q_v) (z_1 \tau ; z_2 \tau ) 
{2 \eta \over \vartheta_2 (z_1 \tau)}
{2 \eta \over \vartheta_2 (z_2 \tau)} \nonumber
\\
&& - (n^2 + \bar n^2) (Q_o - Q_v )(2 z_1 \tau ; 2 z_2 \tau)
{2 \eta \over \vartheta_2 (2z_1 \tau)}
{2 \eta \over \vartheta_2 (2z_2 \tau)} \Biggr\} \ . \label{dannhb}
\end{eqnarray}

The M\"obius amplitude can now 
be recovered, as usual, after a $P$ transformation,
from the transverse amplitudes $\tilde{\cal K}$ and $\tilde{\cal A}$, and
reads
\begin{eqnarray}
{\cal M} &=& - {\textstyle{1\over 4}} \Biggl\{ 
(m+\bar m) (\hat Q_o + \hat Q_v ) (0;0) 2^{(r-4)/2} \sum_\epsilon
\gamma_\epsilon P_1 (B) P_2 (B) \nonumber
\\
&& - (m+ \bar m) (\hat Q_o - \hat Q_v) (0;0) \left( {2 \hat \eta \over
\hat \vartheta_2 (0)}\right)^2 \nonumber
\\
&& - 2^{r/2} (n+\bar n) (\hat Q _o + \hat Q_v ) (2 z_1 \tau ; 2 z_2
\tau ) {2 k_1 \hat \eta \over \hat \vartheta_1 (2 z_1 \tau)}
{2 k_2 \hat \eta \over \hat \vartheta_1 (2 z_2 \tau)} \nonumber
\\
&& - (n+\bar n) (\hat Q_o - \hat Q _v ) (2 z_1 \tau ; 2 z_2 \tau ) 
{2 \hat \eta \over \hat \vartheta_2 (2 z_1 \tau)}
{2 \hat \eta \over \hat \vartheta_2 (2 z_2 \tau)}
\Biggr\} \ ,
\label{dmoebhb}
\end{eqnarray}
where, as in \cite{bps,bsb2}, the $\gamma$'s are signs, required by the
compatibility with the transverse channel, that determine the charge
of the resulting O-planes. 

The R-R tadpoles are modified, and become 
\begin{eqnarray} 
m+\bar m + n + \bar n &=& 2^{5-r/2} \ , \nonumber
\\
k_1 k_2 (n+\bar n) &=& 2^{5-r} \ , \label{tadpoleshb}
\end{eqnarray}
so that 
the ranks of the gauge groups are reduced as usual, albeit 
here in an asymmetrical fashion. 

We can now describe the massless spectrum of these magnetised
orientifolds with a quantised NS-NS background. This clearly
depends on the sign $\gamma$ associated in  ${\cal
M}$ to the massless states, that
determines the type of action (regular or projective) of the
orbifold group on the CP group or, equivalently, the
nature (``real'' or ``complex'') of the CP multiplicities.
 
The more standard choice $\gamma_0 = +1$ requires a
projective $Z_2$ action on the CP labels. Therefore, the resulting 
massless annulus and M\"obius direct-channel amplitudes
\begin{eqnarray} 
{\cal A}_0 & \sim & {\textstyle{1\over 4}} \Bigl\{
4 m\bar m Q_o (0) + 4 n\bar n Q_o (0) + 2 (m^2 + \bar m ^2 ) Q_v (0)
\nonumber
\\
&& + (2\cdot 2^r \cdot k_1 k_2 + 2 \cdot 4 ) (m n + \bar m \bar n ) Q_v
(\zeta \tau) \nonumber
\\
&& + (2 \cdot 2^r \cdot k_1 k_2 - 2 \cdot 4) ( m \bar n + \bar m n )
Q_v (\zeta \tau ) \nonumber
\\
&& + (4\cdot 2^r \cdot k_1 k_2 + 4 ) (n^2 + \bar n ^2 ) Q_v (\zeta \tau)
\Bigr\} \ ,
\label{dannmzhb}
\end{eqnarray}
and 
\begin{equation} 
{\cal M}_0 \sim - {\textstyle{1\over 2}} (m+\bar m) \hat Q_v (0) -
{\textstyle{1\over 2}} (2 \cdot 2^{r/2} \cdot k_1 k_2 + 2) (n+\bar n)
\hat Q_v (\zeta \tau) \ ,
\label{dmoebmzhb}
\end{equation}
involve the ``complex'' multiplicities $m$ and $n$.

Naively, these amplitudes would seem  inconsistent:
as a result of the further multiplicities
related to the rank $r$ of $B_{ab}$, {\it only some} of the string states 
with identical U(1) charges at their ends appear to
contribute to ${\cal M}$. This is actually not the case, and
the solution of the little puzzle follows a pattern that emerged from the
study of SU(2) WZW models \cite{pss}. 
The multiplicities in the annulus count in general different, independent, 
sets of states, that are individually (anti)symmetrised by the 
M\"obius amplitude, so that the corresponding coefficients in ${\cal A}$ and
in ${\cal M}$ need only be equal modulo 2. The
low-lying expansions
\begin{eqnarray}
Q_o (0) &\sim V_4 - 2 C_4 \ \ ;
\qquad Q_v (0) &\sim 4 O_4 - 2 S_4 \ ; \nonumber
\\
Q_o (\zeta\tau ) &\sim {\rm massive}\ ;
\qquad
Q_v (\zeta\tau ) &\sim 2 O_4 - S_4 \ ;
\end{eqnarray}
thus yield the massless spectrum
\begin{eqnarray}
& & ( A + \overline{A},1) + {2 \cdot 2^r \cdot k_1 k_2 + 2 \cdot
4 \over 4} (m,n) + {2 \cdot 2^r \cdot k_1 k_2 - 2 \cdot
4 \over 4} (m,\bar n) \nonumber
\\
& & + \left[ k_1 k_2 \cdot (2^r + 2^{r/2}) + 2 \right] (1, A) +
k_1 k_2 \cdot (2^r - 2^{r/2}) (1, S) \ ,
\end{eqnarray}
with gauge group ${\rm U} (m) \times {\rm U} (n)$, where $S(A)$
denotes the corresponding (anti)-sym\-metric representation. 

Altogether, the tadpole equations admit four inequivalent solutions, and
the corresponding spectra (aside from the universal ${\cal N} =
(1,0)$ gravity multiplet) are summarised in table 1.

\begin{table}
\begin{center}
\begin{tabular}{|c|c|c|c|c|l|} 
\hline
$r$ & $(k_1,k_2)$ & $n^{c}_{T}$ & $n^{c}_{H}$ & $G_{{\rm CP}}$  & 
charged hypermultiplets  
\\
\hline
\hline
2 & (1,1) &  5 & 16 & ${\rm U} (4) \times {\rm U} (4)$ & 
$(6+ \overline{6} ,1)+ 4\,(4,4)$
\\
 & & & & & $+ 8\,(1,6)+ 2\, (1,10)$
\\
\hline
2 & (1,2) & 5 & 16 & ${\rm U} (6) \times {\rm U} (2)$ & 
$(15+ \overline{15} ,1_0)+ 6\,(6,2_+)$
\\
 & & & & & $+ 2\, (6,2_-) + 14\, (1,1_{++})+ 4\, (1,3)$
\\
\hline
2 & (2,2) & 5 & 16 & ${\rm U} (7) \times {\rm U} (1)$ & 
$(21+ \overline{21} ,1_0)+ 10\,(7,1_+)$
\\
 & & & & & $+ 6\,(7,1_-)+ 8\, (1,1_{++})$
\\
\hline
4 & (1,1) & 7 & 14 & ${\rm U} (3) \times {\rm U} (1)$ & 
$(3+ \overline{3} ,1)+ 10\,(3,1_+)$
\\
 & & & & & $+ 6\,(3,1_-) + 12\, (1,1_{++})$
\\
\hline
\end{tabular}
\end{center}
\caption{Massless spectra for $\gamma_0 =+1$.}
\end{table}

As in the non-magnetised case, the choice $\gamma_\epsilon =-1$ in the 
M\"obius amplitude induces a regular action of the $Z_2$ orbifold on
the CP charges \cite{bsb2}. The corresponding multiplicities, now ``real'', 
require also a different embedding of the magnetic U(1)'s, so that
\begin{eqnarray}
m + n + \bar m + \bar n &\to& m_1 + n + \bar n + m_2 \ , \nonumber
\\
m + n - \bar m - \bar n &\to& m_1 + n + \bar n - m_2 \ ,
\end{eqnarray}
and the direct-channel annulus and M\"obius massless 
contributions become
\begin{eqnarray}
{\cal A}_0 & \sim & {\textstyle{1\over 2}} (m_1^2 + m_2^2 ) Q_o (0) + n \bar
n Q_o (0) + m_1 m_2 Q_v (0) \nonumber
\\
& & + \Bigl\{ {\textstyle{1\over 4}} \left[ 2\cdot 2^r \cdot k_1 k_2 -
2\cdot 4 \right] m_1 (n+\bar n) 
\nonumber
\\
& & + {\textstyle{1\over 4}} \left[ 2\cdot 2^r \cdot k_1 k_2 +
2\cdot 4 \right] m_2 (n+\bar n) \Bigr\} Q_v (\zeta\tau ) \nonumber
\\
& &+ {\textstyle{1\over 2}} \left[ 2 \cdot 2^r \cdot k_1 k_2 - 2\right]
(n^2 + \bar n^2 ) Q_v (\zeta \tau ) \ ,
\label{dannhbn}
\end{eqnarray}
and
\begin{equation} 
{\cal M}_0 \sim - {\textstyle{1\over 2}} \left\{ - (m_1 + m_2 ) \hat Q_o (0)
+ \left[ 2\cdot 2^{r/2}\cdot k_1 k_2 + 2 \right] (n+\bar n) \hat Q_v
(\zeta\tau) \right\}  \ .
\label{rmoebhbn}
\end{equation}

For these models with real CP charges, the untwisted tadpole conditions  
\begin{eqnarray}
m_1 + m_2 + n + \bar n &=& 2^{5-r/2} \ , \nonumber
\\
k_1 k_2 ( n + \bar n) &=& 2^{5-r} \ ,
\end{eqnarray}
have to to be supplemented by the twisted tadpole condition
\begin{equation}
m_1 + n + \bar n = m_2 \ .
\end{equation}

A possible solution with $r=2$ and $k_1 = k_2 =1$ is $m_1 =0$,
$m_2 = 8$ and $n=4$, and yields a massless spectrum with 
a gauge group ${\rm USp} (8)
\times {\rm U} (4)$ comprising,
aside from the ${\cal N} =(1,0)$ gravity multiplet, 5 tensor
multiplets, 16 neutral hypermultiplets, and additional
charged hypermultiplets in the representations $4 (8,4) + 6 (1,6)$. 
As in conventional tori \cite{bps} and orbifolds \cite{bsb2}, 
a continuous Wilson line can actually 
connect these two classes of magnetised vacua.
\vskip 8pt
\noindent{\bf Acknowledgements.} We are grateful to I. Antoniadis and
E. Dudas for an enjoyable and instructive collaboration. 
A.S. is grateful to the Organisers for the kind invitation to present
these results. C.A. is supported by 
the ``Marie-Curie'' fellowship HPMF-CT-1999-00256. 
This work was supported in part by the EEC 
contract HPRN-CT-2000-00122 and
in part by the INTAS project 991590. 


\end{document}